\documentclass[conference]{IEEEtran}
\IEEEoverridecommandlockouts
\usepackage{cite}
\usepackage{amsmath,amssymb,amsfonts}
\usepackage{algorithmic}
\usepackage{graphicx}
\usepackage{textcomp}
\usepackage{xcolor}
\def\BibTeX{{\rm B\kern-.05em{\sc i\kern-.025em b}\kern-.08em
    T\kern-.1667em\lower.7ex\hbox{E}\kern-.125emX}}
\begin{document}

\title{FPGA deep learning acceleration based on convolutional neural network*\\
}

\author{\IEEEauthorblockN{1\textsuperscript{st} Xiong Jun}
\IEEEauthorblockA{\textit{Chongqing University)} \\
\textit{School of Microelectronics and Communication Engineering}\\
Chongqing, China \\
923253779@qq.com}

}

\maketitle

\begin{abstract}
In view of the large amount of calculation and long calculation time of convolutional neural network (CNN), this paper proposes a convolutional neural network hardware accelerator based on field programmable logic gate array (FPGA). First, through in-depth analysis of the forward operation principle of the convolutional layer and exploration of the parallelism of the convolutional layer operation, a hardware architecture of input channel parallelism, output channel parallelism and convolution window deep pipeline is designed. Then in the above architecture, a fully parallel multiplication-addition tree module is designed to accelerate the convolution operation and an efficient window buffer module to implement the pipeline operation of the convolution window. The final experimental results show that the energy efficiency ratio of the accelerator proposed in this article reaches 32.73 GOPS/W, which is 34\% higher than the existing solution, and the performance reaches 317.86 GOPS.
\end{abstract}


\section{Introduction}
In recent years, Convolutional Neural Network (CNN) has been widely used in the field of artificial intelligence \cite{b1}. However, while CNN is constantly approaching the limit of task accuracy, its network depth and the number of parameters are also growing rapidly, which consumes more and more computing resources and memory resources \cite{b2}.

Currently, there are three main platforms for accelerating convolutional neural networks: Graphics Processing Unit (GPU). Its software programmable and multi-CUDA architecture are very suitable for accelerating convolutional neural networks, but its significant power consumption makes it difficult Integrated into embedded platforms with limited power consumption \cite{b3}; Application-Specific Integrated Circuit (ASIC), which has the characteristics of high performance and low power consumption, but has a long design cycle and high manufacturing costs \cite{b4}; Field-Programmable Gate Array (FPGA), with the characteristics of low power consumption and high flexibility, has become the most popular platform for studying convolutional neural network hardware acceleration.

The research of FPGA-based convolutional neural network accelerator design mainly focuses on parallel computing and memory bandwidth optimization. In terms of parallel computing, Motamedi et al. \cite{b5} summarized the convolutional layer parallel computing optimization method in detail, and proposed three parallel computing methods: parallel between outputs, parallel between convolution kernels, and parallel within convolution kernels. If the area, bandwidth, and on-chip storage of the FPGA are not limited, theoretically all the above methods can be used to speed up the neural network to the maximum, but in practice it is impossible. Therefore, the challenge is to study the best combination of multiple parallel mechanisms. In terms of memory bandwidth optimization, Zhang et al. \cite{b6} first store the input feature map and weight data in BRAM, and then read the data from it for convolution operation. Although the weight data is reused, the input feature map is not used. The convolution window is reused, and the memory utilization is not high. In addition, all the above designs use a large number of classic addition tree units in parallel computing, which requires a large bandwidth and takes up a lot of resources.

In order to solve the above problems, this paper has completed the following research work:
\begin{itemize}
    \item Analyze the feasibility of parallel acceleration of the convolutional layer, and propose three parallel computing methods: intra-convolution, input channel parallel, and output channel parallel; 
\item A new addition tree design method is proposed, a fully parallel multiplication-addition tree module is designed, which reduces computing resources and memory resources and maintains the same computing performance;
\item Pipeline operation is adopted for the convolution window, and the designed efficient window buffer module generates a convolution window every clock cycle, thereby improving the performance of pipeline;
\item A convolutional layer hardware accelerator is designed, which performs pipeline operations on input channel parallelism, output channel parallelism, and convolution window, which greatly improves computing performance.
\end{itemize}

\section{Related Work}
\begin{figure}[htbp]
\centerline{\includegraphics[width = 3in]{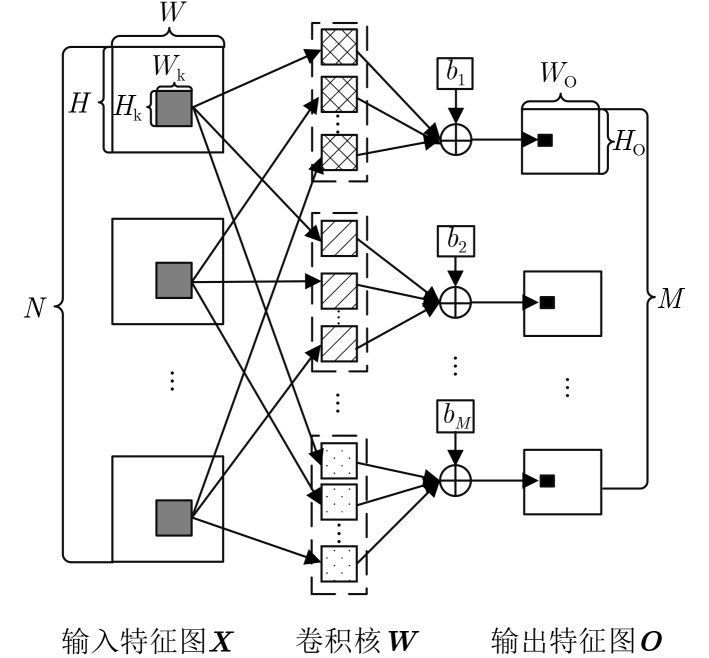}}
\caption{Convolutional layer operation process.}
\label{fig1}
\end{figure}

The calculation in the convolutional neural network is mainly concentrated in the convolutional layer \cite{b7}, Fig.\ref{fig1} shows the overall process of the convolutional layer operation.

The input feature map $X$ in Fig.\ref{fig1} is a three-dimensional matrix whose shape is $[N,H,W] $, where $N$ is the number of channels of the input feature map, and $H$ and $W$ are the height and width of the input feature map, respectively. The convolution kernel $W$ is a 4-dimensional matrix whose shape is $[M,N,H_k,W_k]$, where $M$ is the number of convolution kernels, $N$ is the number of channels of the convolution kernel, that is, the number of channels of the input feature map, and the height and width of the $H_k$ and $W_k$ convolution kernels . The vertical and horizontal steps of the convolution kernel are $H_s$ and $W_s$ respectively. The offset is a 1-dimensional vector of length $M$, denoted as $b$. The output feature map $O$ is a 3-dimensional matrix whose shape is $[M,H_o,W_o]$, where $M$ is the number of channels of the output feature map, that is, the number of convolution kernels, and $H_o$ and $W_o$ are the height and width of the output feature map, respectively. The relation is
\begin{equation}\label{label1}
  H_0=\lfloor \frac{H-H_k}{H_s} \rfloor+1
\end{equation}
\begin{equation}\label{label2}
  W_0=\lfloor \frac{W-W_k}{W_s} \rfloor+1
\end{equation}

Convolution output feature map $m$ channel $i^{\prime}$ row $j^{\prime}$ column data $O_{mi^{\prime}j^{\prime}}$ satisfy

\begin{equation}\label{label3}
  O_{mi^{{\prime}}j^{{\prime}}} = \sum\limits_{n=1}^N{\sum\limits_{i=1}^{H_k}{\sum\limits_{j=1}^{W_k}{X_{n(i+(i^{\prime}-1)H_s)(j+(j^{\prime}-1)W_s)}}}}
\end{equation}

If the channel size of a certain convolution kernel is $3 \times 3$, the channel size of the input feature map is $5 \times 5$, and the step size in both the vertical and horizontal directions is 2, then it obtains the output channel component of the size $2 \times 2$.

\section{Math}
\subsection{Feasibility analysis of convolutional layer parallel acceleration}

\begin{figure}[htbp]
\centerline{\includegraphics[width = 3in]{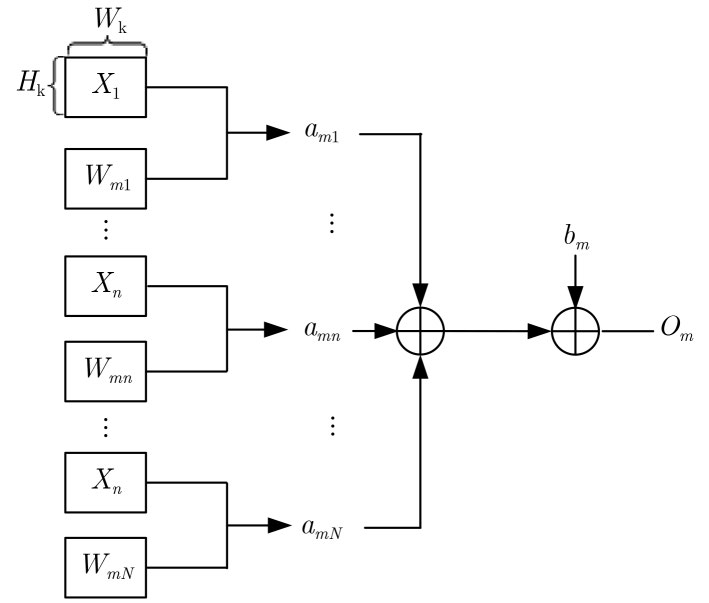}}
\caption{Parallel calculation of convolution windows for N input channels.}
\label{fig2}
\end{figure}

Since each convolution window has the same operation method, for the convolution window located in the upper left corner of the input feature map, namely $i^{\prime}=1,j^{\prime}=1$, equation \eqref{label3} can be rewritten as
\begin{equation}\label{label4}
  O_m = \sum\limits_{n=1}^N{\sum\limits_{i=1}^{H_k}{\sum\limits_{j=1}^{W_k}{X_{nij}W_{mnij}+b_m}}}
\end{equation}
Among them, $m \in [1,M]$, the output value corresponding to the convolution window has $M$, respectively $O_1,O_2, \cdots ,O_M$. Fig.\ref{fig2} is a schematic diagram of the above calculation process.

According to equation \eqref{label4}, let
\begin{equation}\label{label5}
  a_{mn} = \sum\limits_{i=1}^{H_k}{\sum\limits_{j=1}^{W_k}{X_{nij}W_{mnij}}}
\end{equation}

Then there is
\begin{equation}\label{label6}
  O_m = \sum\limits_{n=1}^N{a_{mn}+b_m}
\end{equation}

To calculate the values of $O_1,O_2, \cdots ,O_M$ and $M$, the following two methods can be used:

(1) From equation \eqref{label6}, we can first calculate $a_{m1},a_{m2}, \cdots ,a_{mN}$, then sum the numbers of $N$ and add $b_m$ to get $O_m$. Let $m$ take $[1,M]$, then $O_1,O_2, \cdots ,O_M$ can be calculated.

(2) It can be seen from equation \eqref{label6},
\begin{equation}
\begin{split}
  \label{label7}
O=\begin{bmatrix}O_1\\O_2\\\vdots\\O_M\end{bmatrix}=\begin{bmatrix}a_{11}\\a_{21}\\\vdots\\a_{M1}\end{bmatrix}+\begin{bmatrix}a_{12}\\a_{22}\\\vdots\\a_{M2}\end{bmatrix}+\cdots+\\\begin{bmatrix}a_{1N}\\a_{2N}\\\vdots\\a_{MN}\end{bmatrix}+\begin{bmatrix}b_1\\b_2\\\vdots\\b_M\end{bmatrix}=\hat{O_1}+\hat{O_2}+ \cdots+\hat{O_N}+b  
\end{split}
\end{equation}
Among them, $\hat{O_n}$ is the $n$ component of the convolution output $O$, which satisfies
\begin{equation}
\label{label8}
\hat{O_n}=\begin{bmatrix}a_{1n}\\a_{2n}\\\vdots\\a_{Mn}\end{bmatrix} 
\end{equation}
Among them, $n \in [1,M]$.

\begin{figure}[htbp]
\centerline{\includegraphics[width = 3in]{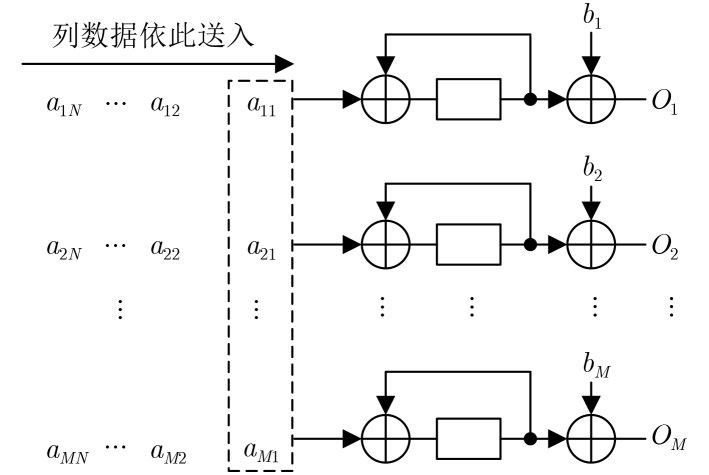}}
\caption{Accumulator parallel operation.}
\label{fig3}
\end{figure}

As shown in  Fig.\ref{fig3}, using $M$ accumulators, first find $a_{11},a_{21}, \cdots ,a_{M1}$ and store them in $M$ registers respectively, then find and accumulate $a_{12},a_{22}, \cdots ,a_{M2}$, find and accumulate $a_{1N},a_{2N}, \cdots ,a_{MN}$ at the $N$ time, and finally add offset $b_1,b_2, \cdots ,b_M$, you can find $O_1,O_2, \cdots ,O_M$.

It can be seen that there are 3 parts that can be calculated in parallel in the convolutional layer:

(1) In equation \eqref{label5}, $H_kW_k$ multiplications can be calculated in parallel.
Parallel is located inside the convolution kernel, which is called intra-convolution parallel;

(2) In equation \eqref{label6}, when calculating the $m$ convolution output $O_m$, the convolution of the corresponding $N$ input channels can be calculated in parallel, and the $a_{m1},a_{m2}, \cdots ,a_{mN}$ and $N$ intermediate results can be obtained. Finally, the $N$ values are summed to get $O_m$. This is the input channel parallel;

(3) In equation \eqref{label7}, when calculating the $n$ component of the convolution output $O$, the convolution of the $n$ input channel and the corresponding channel in the $M$ convolution kernel can be calculated in parallel, which is the parallel output of the output channels.
\subsection{Basic module design}
\subsubsection{Fully parallel multiplication-addition tree module}

It can be seen from equation \eqref{label3} that each convolution window of each output channel of the convolution operation contains $N \times H_k \times W_k$ multiplications, so the entire convolution layer has $M \times G \times N \times H_k \times W_k$ multiplications. The corresponding summation symbol in equation \eqref{label3} is the addition calculation. And each window of each output channel needs to sum the number of$N \times H_k \times W_k+1$, so the entire convolutional layer needs to sum the number of $M \times G \times (N \times H_k \times W_k+1)$. Therefore, such multiplication and addition operations must be optimized in parallel.
The shape of the convolution kernel is generally square. Assuming the size of the convolution kernel is $W_k=H_k=K$, the convolution output $y$ can be obtained by equation \eqref{label3}
\begin{equation}
\label{label9}
 y=\sum_{i=1}^K{\sum_{j=1}^K{x_{ij}w_{ij}}}
\end{equation}

\begin{figure}[htbp]
\centerline{\includegraphics[width = 3.5in]{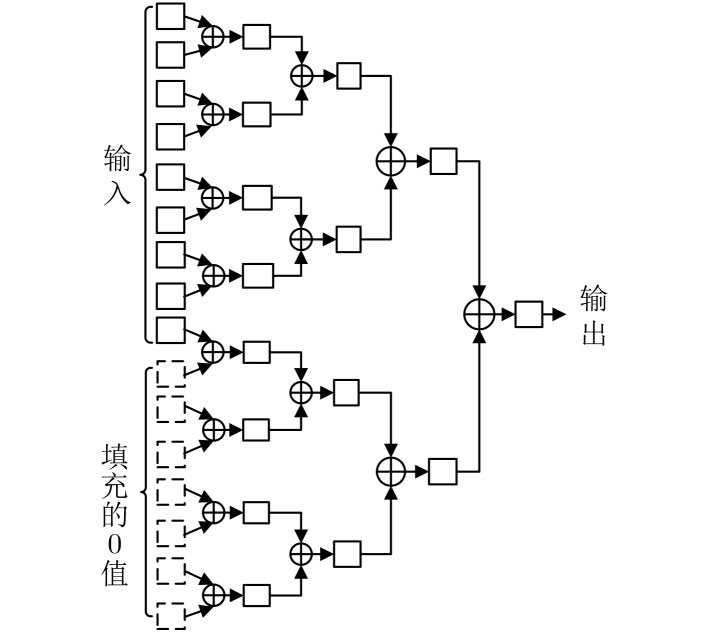}}
\caption{Classic addition tree.}
\label{fig4}
\end{figure}

It can be seen from equation \eqref{label9} that the operation includes $K^2$ multiplication operations and $K^2$ number addition operations. For $K^2$ multiplication operations, $K^2$ multipliers are used for full parallel calculation; for the addition of $K^2$ numbers, in hardware acceleration design, the classic addition tree is generally used to achieve \cite{b8}, this addition tree first passes the input number The way of filling 0 is expanded from $K^2$ to $2^{\lceil \log_2({K^2})\rceil}$, and then the sum of every two numbers is used as the input of the second layer. In this way, it is accumulated step by step until the final layer gets the sum. Fig.\ref{fig4} shows the structure of this design for $K=3$.

Although this addition tree structure greatly improves the parallelism of addition, it also has the following two disadvantages:

(1) Consume too many hardware resources
For the addition of the number of $\eta$, the number of adders required by this addition tree is $f_1(\eta)$, the number of registers required is $g_1(\eta)$, and the required clock cycle $h_1(\eta)$ is respectively

For example, for the addition operation of 144 numbers and 256 numbers, this kind of addition tree requires that the number of adders is 255, the number of registers is 511, and the clock cycle is 8. It can be seen that in the above two cases, although the number of addition inputs is different, they consume the same number of adders, registers, and clock cycles. Obviously, this method wastes computing resources and memory resources when the number of addition inputs is slightly larger than the power of two.

(2) Large bandwidth demand

When $K=12$, the number of data that the classical addition tree needs to be calculated at the same time is increased from 144 to 256, and the bandwidth requirement is almost doubled.

In view of the above problems, the improvement of the additive tree designed in this paper is as follows:

(1) If the number of inputs in the current layer is even, it can be the same as the classic addition tree, adding every two numbers;

(2) If the number of inputs of the current layer is odd, first use the method (1) for parallel calculation for the even number, and finally output the remaining addend directly to the next layer.

\begin{figure}[htbp]
\centerline{\includegraphics[width = 3.5in]{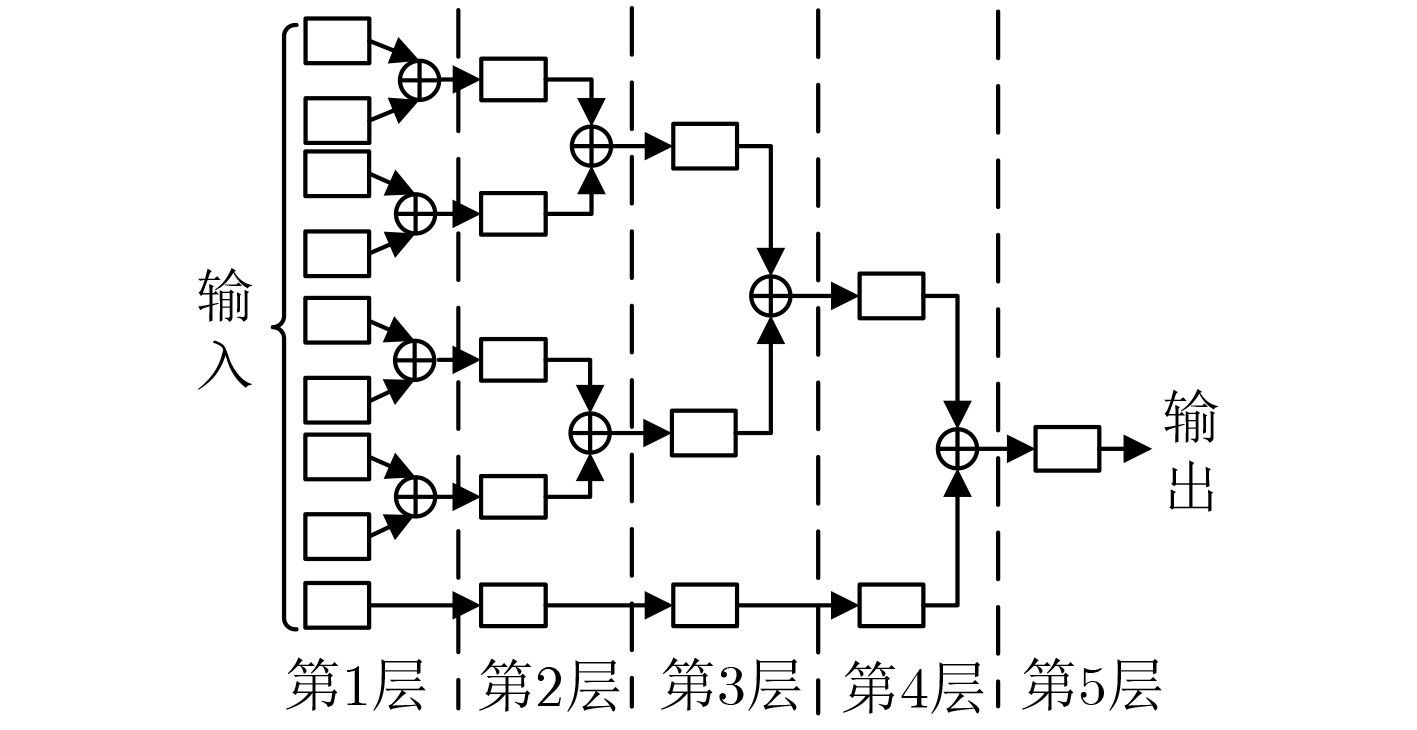}}
\caption{Addition tree.}
\label{fig5}
\end{figure}

This kind of addition tree does not need to add 0, that is, additional memory, and does not need an additional adder, and the required clock cycle is the same as the classic method. For the addition of 9 numbers, the addition tree designed in this article is shown in Fig.\ref{fig5}.

It can be seen from Fig.\ref{fig5} that for the addition of 9 numbers, the addition tree designed in this article requires 8 adders, 20 registers, and 4 clock cycles. The corresponding classic addition tree requires 15 adders, 31 registers, and a calculation clock cycle of 4.

Suppose the number of inputs in the additive tree is $\eta$, then the number of inputs in the next layer is $\lceil \eta /2\rceil$, and so on.

The operation steps of the multiplication-addition tree module are:

(1) Store the input feature map matrix and the convolution kernel weight matrix in the buffer area, which are called input buffer and weight buffer respectively;

(2) Read data from the input buffer and weight buffer, and use $K^2$ multipliers to calculate the $K^2$ multiplications in equation \eqref{label9} in parallel to obtain $K^2$ intermediate results;

(3) Finally, use the method of designing the addition tree in this paper to build an addition tree with the number of addition inputs $K^2$. The input of the addition tree is the $K^2$ intermediate results obtained in (2), and the output is the final result of the multiplication-addition module.

\subsubsection{Efficient window cache module}
According to equation \eqref{label3}, it can be seen that when the convolution layer is calculated, many convolution windows are generated. From equation \eqref{label1} and equation \eqref{label2}, it can be seen that the number of convolution windows generated is $G=H_0W_0$. Suppose the size of the convolution kernel is $W_k=H_k=K$, and the shape of a channel matrix $x$ of the input feature map is $[H,K]$, let
\begin{equation}
\label{label10}
x=\begin{bmatrix}x_{11}&x_{12}&\cdots&x_{1W}\\x_{21}&x_{22}&\cdots&x_{2W}\\\vdots&\vdots&\ddots&\vdots\\x_{H1}&x_{H2}&\cdots&x_{HW}\\\end{bmatrix}
\end{equation}

For the convenience of explanation, the subscript of matrix $x$ becomes continuous, that is, the subscript of the element in row $i$ and column $j$ is $j+(i-1)$, then matrix $x$ is re-expressed as
\begin{equation}
\label{label11}
x=\begin{bmatrix}x_{1}&x_{2}&\cdots&x_{W}\\x_{W+1}&x_{W+2}&\cdots&x_{2\times W}\\\vdots&\vdots&\ddots&\vdots\\x_{(H-1)W+1}&x_{(H-1)W+2}&\cdots&x_{H\times W}\\\end{bmatrix}
\end{equation}

\begin{figure}[htbp]
\centerline{\includegraphics[width = 3.5in]{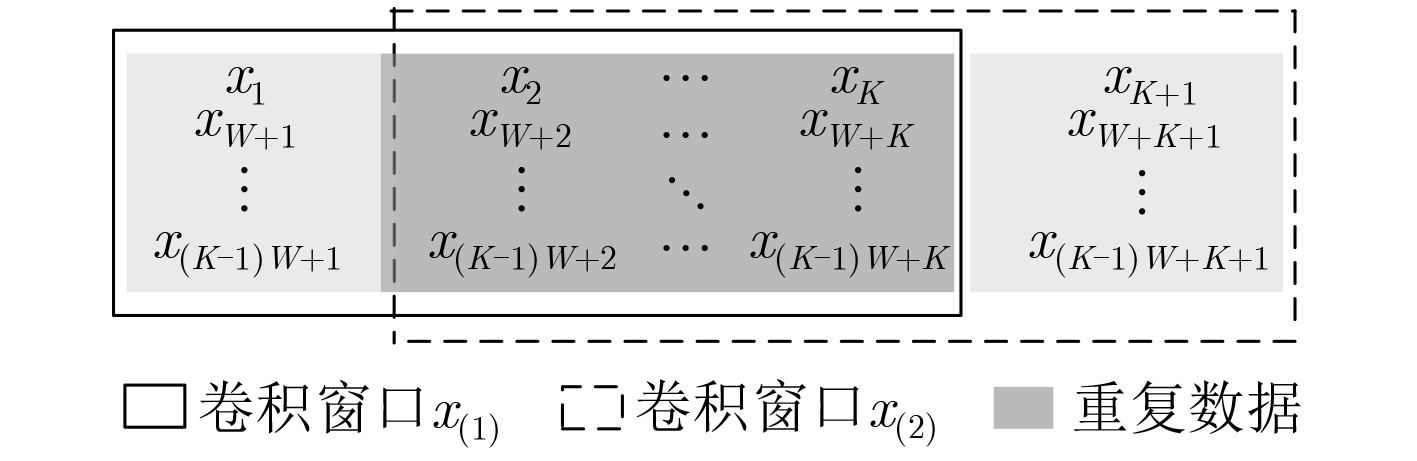}}
\caption{Convolution window data reuse.}
\label{fig6}
\end{figure}

Definition $x_{(i)}$ represents the $i$ convolution window, where $i \in [1,G]$. As shown in Fig.\ref{fig6}, the data in the second to the $K$ column of  $x_{(1)}$ and the first to the $K-1$ column of $x_{(2)}$ are exactly the same. The first convolution window and the second convolution window all have $2 \times K \times K$ data, and the shared data has  $2 \times K \times (K-1)$ so the data repetition ratio is $(K-1)/K$, and when the value of $K$ is greater, the data between the two windows is repeated The proportion is also larger.

If hardware parallelism is used for all convolution windows, too much computing resources will be consumed. Therefore, this article considers pipelining the convolution window, which can greatly reduce storage resources and computing resources. These convolution windows use the same circuit structure, and input different convolution window data at different times to obtain convolution operation results. In order to pipelining the convolution window, this paper designs a window buffer module, which consists of two 2-dimensional registers, one of which is the window buffer register WINDOW\_BUFFER, the other is the shift register SHITF\_BUFFER, as shown in Figure 7 for the window Schematic diagram of the cache structure.

As can be seen from Fig.\ref{fig7}, the storage size of WINDOW\_BUFFER is $K \times K$, and the storage size of SHIFT\_BUFFER is $(K-1) \times (W-K)$, and the following 5 steps are performed in parallel in each clock cycle:
\begin{figure}[htbp]
\centerline{\includegraphics[width = 3.5in]{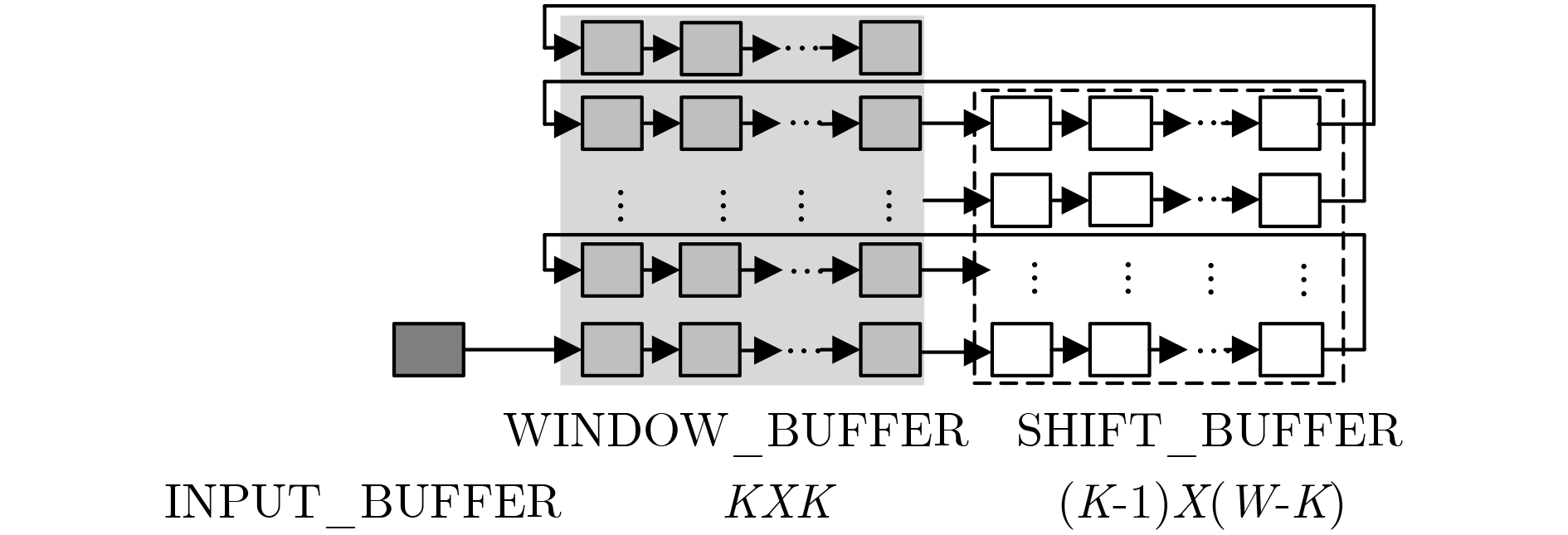}}
\caption{Window cache structure.}
\label{fig7}
\end{figure}

(1) Input a piece of data from the input buffer INPUT\_BUFFER to the first column of row $K$ of WINDOW\_BUFFER;

(2) WINDOW\_BUFFER performs right shift operation every time;

(3) The data in the second column to the $K$ row of WINDOW\_BUFFER is assigned to all the rows of the first column of SHIFT\_BUFFER;

(4) SHIFT\_BUFFER right shift operation for each line;

(5) The data in the last column of SHIFT\_BUFFER is assigned to the first column from row 1 to row $K-1$ of WINDOW\_BUFFER.

\begin{figure}[htbp]
\centerline{\includegraphics[width = 3.5in]{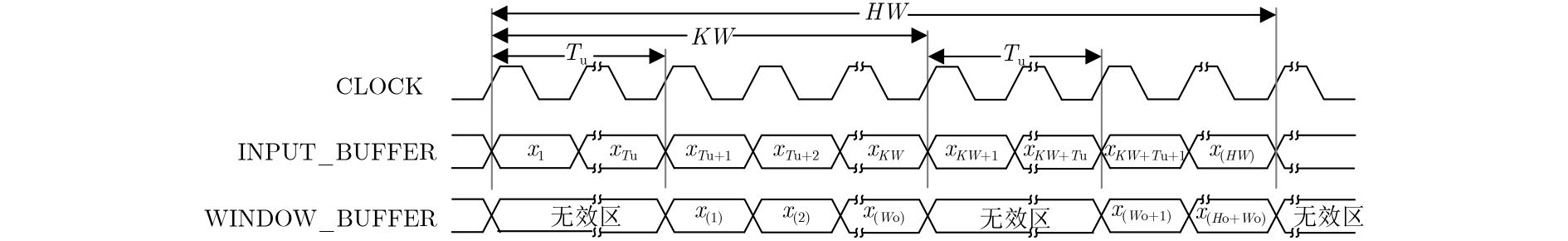}}
\caption{Window cache structure.}
\label{fig8}
\end{figure}

Fig.\ref{fig8} shows the timing diagram of the above steps. The data in the WINDOW\_BUFFER register in the previous $(K-1) \times W+K-1$ clock cycle is invalid, which is called the invalid area, which defines the invalid clock cycle $T_u=(K-1) \times W+K-1$; at the $T_u+1$ clock cycle, the data stored in WINDOW\_BUFFER is $x_{(1)}$ ; After another clock cycle, the data stored in WINDOW\_BUFFER is $x_{(2)}$; and so on, when the $K \times W$ clock cycle, the data stored in WINDOW\_BUFFER is $x_{(W_0)}$; finally, at the $H \times W$ clock cycle, WINDOW\_BUFFER is stored The data is the last convolution window, which is$x_{(H_0W_0)}$.

From the Fig.\ref{fig8} and the previous analysis, we can see that the window buffer module designed in this article can generate a convolution window for subsequent operations in each clock cycle, which is the most efficient window pipeline method.
\section{Experiment}
\subsection{Experiment environment}

In this experiment, a multi-layer convolutional neural network was trained on the GPU platform to classify the MNIST data set \cite{b9}, and then accelerated forward inference was implemented on the FPGA platform. The FPGA platform used is the Cyclone V (5CGXFC9D6F27C7) development board developed by ALTERA, which contains DDR3 SDRAM with a maximum bandwidth of 2GB/s and a capacity of 1GB. And through the PLL frequency multiplication, the FPGA operating frequency is increased to 100 MHz. The used neural network structure parameters are shown in Tab \ref{tab1}.
\begin{table*}[htp]
\begin{center}
\caption{Convolutional neural network structure parameters}
\begin{tabular}{lcl}
\hline
\multicolumn{1}{c}{\textbf{Layer name}} & \textbf{Layer structure}                                                   & \multicolumn{1}{c}{\textbf{Parameter amount}} \\ \hline
Convolutional layer 1                   & Convolution kernel size 3×3, number of convolution kernel 15, step size 1  & 150                                           \\
Active layer 1                          & No                                                                         & 0                                             \\
Pooling layer 1                         & Pooling size 2×2, step size 2                                              & 0                                             \\
Convolutional layer 2                   & Convolution kernel size 6×6, number of convolution kernels 20, step size 1 & 10820                                         \\
Active layer 2                          & No                                                                         & 0                                             \\
Pooling layer 2                         & Pooling size 2×2, step size 2                                              & 0                                             \\
Fully connected layer                   & Number of output neurons 10                                                & 3210                                          \\ \hline
\end{tabular}
\label{tab1}
\end{center}
\end{table*}

In addition, comparative tests were conducted on CPU and GPU. The CPU and GPU software codes all use the TensorFlow\cite{b10} deep learning framework, and the underlying operation code is C++. The CPU uses Intel(R) CoreTM i7-7700, the main frequency is 3.6GHz; the GPU uses NVIDIA GTX1080Ti, the peak double-precision floating-point computing performance is 5.75TFLOPS, the peak single-precision floating-point computing performance is 11.5TFLOPS, and the memory capacity It is 11 GB, the memory bandwidth is 484 GB/s, the number of CUDA cores is 3584, and the core frequency is 1.582 GHz.
\subsection{Experimental results}

The resource consumption on FPGA circuit board is shown in Tab \ref{tab2}. The DSP resource consumption in Tab \ref{tab2} is more, mainly because more multipliers are used in the convolution operation.
\begin{table}[htbp]
\begin{center}
\caption{FPGA resource consumption}
\begin{tabular}{lcl}
\hline
\multicolumn{1}{c}{\textbf{}} & \textbf{Resources} & \multicolumn{1}{c}{\textbf{Proportion(\%)}} \\ \hline
ALMs                          & 89423/113560       & 79                                          \\
Block Memory                  & 730151/12492800    & 6                                           \\
DSPs                          & 342/342            & 100                                         \\ \hline
\end{tabular}
\label{tab2}
\end{center}
\end{table}

\begin{figure}[htbp]
\centerline{\includegraphics[width = 3.5in]{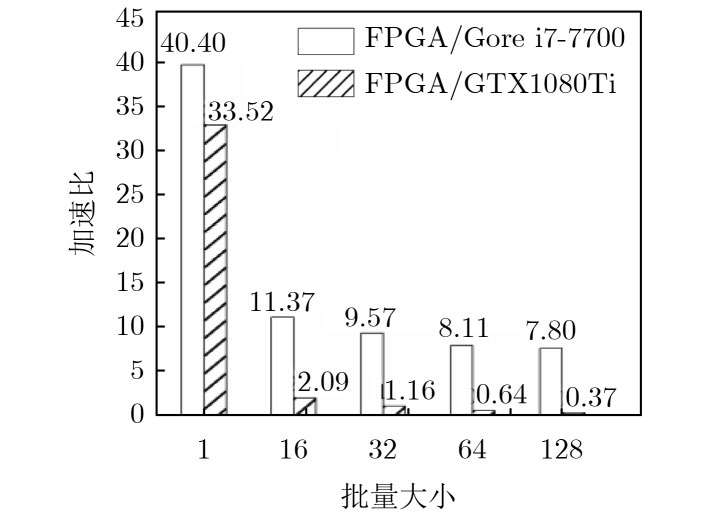}}
\caption{FPGA, CPU, GPU performance comparison.}
\label{fig9}
\end{figure}

As shown in Fig. \ref{fig9} when only one picture is used for forward inference, that is, when the batch size is 1, FPGA is about 40.4 times faster than CPU and about 33.52 times faster than GPU. As the batch size increases, FPGAs are still more than 7.8 times faster than CPUs. But when the batch size is increased to more than 64, the advantage of GPU is highlighted, because GPU computing is characterized by large bandwidth, which can meet the simultaneous parallel computing of multiple pictures. However, in the task of real-time video detection on embedded platforms, the processing of pictures one by one, that is, when the batch size is 1, the advantage of FPGA is very obvious, and the power consumption is much lower than that of GPU, which is more suitable for practical applications. use.

Tab \ref{tab3} compares the results with other convolutional neural network FPGA implementations. Since the literature uses different FPGA devices and different network structures, if only performance is used as a reference index, it will lack fairness, so it also increases the power efficiency. Parameters for comprehensive comparison of acceleration effects.
\begin{table*}[htbp]
\begin{center}
\caption{Compared with literature FPGA hardware acceleration}
\begin{tabular}{lllll}
\hline
\multicolumn{1}{c}{\textbf{}}    & \multicolumn{1}{c}{\textbf{Document {\cite{b7}}}} & \multicolumn{1}{c}{\textbf{Document {\cite{b11}}}} & \textbf{Document {\cite{b12}}} & \textbf{The method of this paper} \\ \hline
FPGA                             & \multicolumn{1}{c}{ZynqXC7Z045}               & ZynqXC7Z045                                    & Virtex-7 VX690T            & Cyclone V 5CGXF                   \\
Frequency (MHz)                  & \multicolumn{1}{c}{150}                       & 100                                            & 150                        & 100                               \\
DSP resources                    & \multicolumn{1}{c}{780}                       & 824                                            & 1376                       & 342                               \\
Quantitative strategy            & 16 bit fixed                                  & 16 bit fixed                                   & 16 bit fixed               & 16 bit fixed                      \\
Power consumption (W)            & 9.630                                         & 9.400                                          & 25.000                     & 9.711                             \\
Performance (GOPS)               & 136.97                                        & 229.50                                         & 570.00                     & 317.86                            \\
Energy efficiency ratio (GOPS/W) & 14.22                                         & 24.42                                          & 22.80                      & 32.73                             \\ \hline
\end{tabular}
\label{tab3}
\end{center}
\end{table*}

It can be seen from Tab \ref{tab3} that the energy efficiency ratio of the accelerator designed in this paper reaches 32.73 GOPS/W, which is better than the other three schemes, and is 34\% higher than the best implementation method in the existing literature \cite{b11}. In addition, its acceleration performance has reached 317.86 GOPS. Compared with literature \cite{b7} and literature \cite{b11}, the acceleration effect is obvious, but there are still some gaps in performance compared with literature \cite{b12}. The reason is that the FPGA platform used in this article has a DSP resource ratio Other platforms are much less, and the number of DSPs is one of the key factors to improve accelerator performance. Although limited by resources, the test platform did not give full play to the optimal performance of the accelerator designed in this paper. However, the existing results have shown that the accelerator implemented in this paper has a high energy efficiency ratio and good acceleration performance.

\section{Conclusion}
This article deeply analyzes the principle and feasibility of parallel acceleration of convolutional layers
Performance, redesigned the addition tree module, combined with the multiplication design
A universal fully parallel multiplication-addition tree module, and then designed
The efficient window cache module pipelines the convolution window,
And designed an acceleration scheme with parallel input channels and parallel output channels
Accelerate the convolution operation and finally build a complete forward inference
Convolutional neural network FPGA accelerator. Through experimental analysis, this article
The designed accelerator energy efficiency ratio reaches 32.73 GOPS/W, which is higher than the existing
The solution is 34\% higher. Show that the convolutional neural network implemented in this article
Network hardware accelerator has a high energy efficiency ratio, which is very suitable for
Embedded platform.

\vspace{12pt}

\end{document}